\documentstyle[11pt,twoside,epsfig,jltp]{article}
\runninghead{D. C. Dixon, C. P. Heij, P. Hadley, and J. E. Mooij }{Enhancement of Josephson quasiparticle current...}

\begin{document}

\title{Enhancement of Josephson quasiparticle current in coupled superconducting
single-electron transistors}
\author{D. C. Dixon, C. P. Heij, P. Hadley, and J. E. Mooij \address{Department of
Applied Physics and DIMES, TU-Delft, Delft, The Netherlands}}

\begin{abstract}
The Josephson quasiparticle (JQP) cycle in a voltage-biased 
superconducting single-electron
transistor (SSET) combines coherent Cooper pair tunneling with incoherent
quasiparticle decay.  We have measured the influence of current flow through an 
independently-biased SSET on the JQP cycle when the two SSET's have a 
strong mutual capacitive coupling.  We find, among other effects, 
that the JQP current in one SSET is enhanced by the presence 
of a quasiparticle current in the other SSET.  A simplified model
of the coupled-SSET system is presented which reproduces the 
enhancement effect.

PACS numbers: 73.23.Hk, 74.50.+r, 85.30.Wx, 85.25.Na 
\end{abstract}

\maketitle
\section{INTRODUCTION}
Superconducting single-electron transistors (SSET's) are small islands of
superconducting material isolated from an external circuit by tunnel
barriers (Josephson junctions). The normal tunnel barrier resistances 
($R>h/4e^{2}\sim $6.5 k$\Omega$) are sufficient to constrain
the excess charge on the island to integer multiples of $e$. 
At equilibrium, adding an electron to an electrically neutral
island costs a charging energy $E_{c}=e^{2}/2C$, where $C$ is the
island's total capacitance.  This "Coulomb blockade" may be lifted by
applying a gate voltage to the SSET, or it may be surmounted by applying
a sufficient source-drain bias voltage.

At zero voltage bias, a supercurrent of Cooper pairs may flow coherently through the SSET,
while for a large voltage bias ($|eV|>4\Delta$, where $\Delta$ is
the superconducting gap), the current is dominated
by successive charging and discharging of the island by quasiparticles.
Within a range of moderate bias ($2\Delta +E_{c}<|eV|<2\Delta +3E_{c}$),
current flows via a hybrid transport mechanism termed the ''Josephson
quasiparticle cycle'' (JQP)$^{1,2}$. In each turn of the cycle, a
Cooper pair (CP) resonantly tunnels into the SSET through one junction, accompanied by 
a quasiparticle (QP) tunneling event through the
other junction, leaving one extra QP on the island.  This extra QP 
then also tunnels through the second junction, and the cycle starts anew.  

\begin{figure}
\epsfig{file=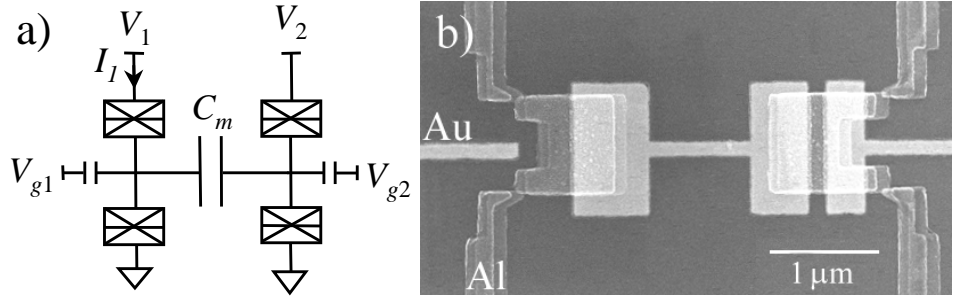, width=5in, clip=true}
\caption{a) Schematic circuit diagram
of the device. Each SSET is bounded by two Josephson junctions, a gate
capacitance, and a large mutual coupling capacitance $C_{m}$. (b) SEM micrograph of
the device. The two Al/AlO$_{x}$ islands are coupled to the leads at the
corners by tunnel barriers, and to each other by an underlying
dumbbell-shaped strip of Au, acting as an overlap capacitor (actually two
overlap capacitors in series). The gate electrodes, seen to the left and
right, are also formed in the Au layer.}
\end{figure}

In this paper we show some intriguing measurements of the JQP current flowing through
a SSET that is strongly coupled to a nearby, independently voltage-biased
SSET. The two islands are coupled by a large capacitance $C_{m}$ supplied by
an overlap capacitor instead of a tunnel junction, so there is no 
Josephson coupling between the islands themselves.  The charging energy 
associated with this mutual capacitance is given by:
\begin{equation}
E_{m} = {{e^2 C_{m}}\over{C_{1}C_{2} - C_{m}^2}} < E_c
\end{equation}
where $C_{i}$ is the total capacitance of island $i$ (including $C_{m}$).  
The strong capacitive coupling
ensures that the charge of one island influences the QP and CP tunneling
rates for the other island, since $E_{m} >> kT$.  

The islands were formed by the standard procedure of double angle evaporation
of Al, with an oxidation step to form AlO$_{x}$ tunnel barriers between each
island and its leads (Fig. 1(b)). The gate 
electrodes and the central metal strip
coupling the two islands were created in an underlying Au layer, with an
intermediate insulation layer of SiO (32 nm). From normal-state measurements 
\cite{heij99}, the junction capacitances were determined to be $\sim$ 0.3 fF, 
while $C_{m}\sim $ 0.6 fF. The series resistances of the SSET's were
7 and 13 M$\Omega$, so the Josephson energies of the tunnel junctions are
expected to be very small ($E_{J} < $ 0.1 $\mu $eV) compared to the charging
energy ($E_{c} \sim$ 80 $\mu$eV) and $E_m \sim$ 40 $\mu$eV.

Measurements of the device were carried out at low electron temperature (27
mK) in a helium dilution refrigerator. The source-drain voltage biases for
the two SSET's ($V_{1},V_{2}$) were applied asymmetrically, with one lead
from each SSET tied to ground (Fig. 1(a)). 

\section{NORMAL STATE MEASUREMENTS}

Before discussing the JQP experiments, we will discuss some relevant properties 
of the system in the normal state.  After applying a 1 T magnetic
field to suppress superconductivity, we measured low bias ($V_{1}$ = 20 $\mu$V) Coulomb oscillations in the left SET for various values of $V_{2}$.  The results are plotted
in Figure 2.  We observe that when $V_{2}$ is sufficient to allow single-electron
transport through the right SET, the Coulomb peaks spread, develop small sidepeaks, 
and diminish in amplitude.  These measurements agree with our
simulations based on the orthodox theory of single-electron tunneling
\cite{orthodox}.

The sidepeaks are due to the discrete charging of the right SET.  The presence of an extra electron induces a charge of $\sim$ 0.25$e$ on the left SET via $C_{m}$; this shifts the Coulomb peak to a different $V_{g1}$, resulting
in the extra peaks marked by arrows in Fig. 2(b).  Each peak can thus be indexed by the charge of the right SET ($n_2$). 

Note that the peaks (and sidepeaks) are not strictly $e$-periodic ($\Delta
V_{g1}=e/C_{g1}\sim $ 4 mV), since $V_{g1}$ has a cross-capacitance to 
the right SET.  This effect can be 
cancelled by countersweeping $V_{g2}$ in proportion to $V_{g1}$.  

\begin{figure}
\epsfig{file=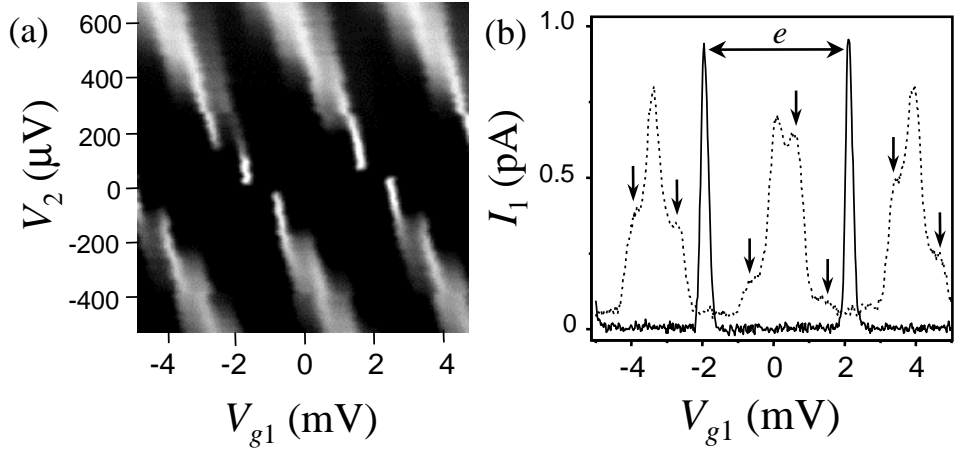, width=5in, clip=true}
\caption{(a) Current $I_1$ through the left SET (normal state), plotted in grayscale as a
function of $V_{g1}$ and $V_{2}$ (black indicates no current, while white corresponds to 1 pA). (b) Two line traces selected from the data in
(a) (solid line: $V_2$ = 0, dashed line: $V_2$ = 400 $\mu$V)}
\label{fig2}
\end{figure}

\begin{figure}
\epsfig{file=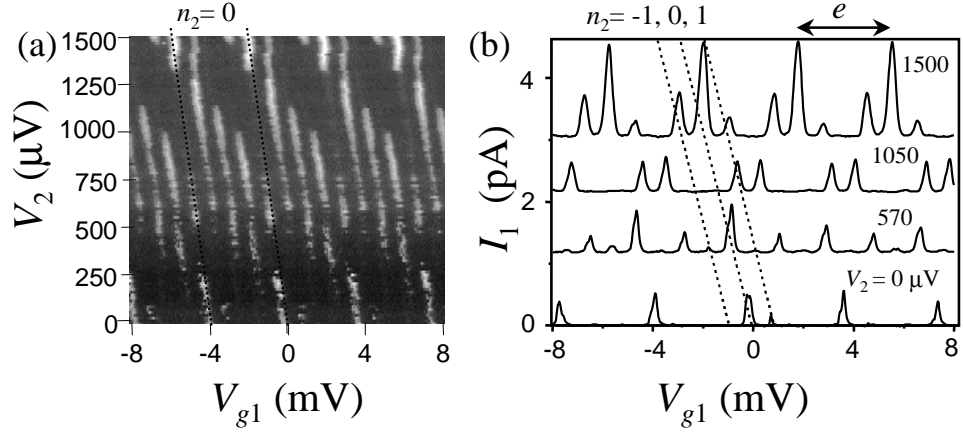, width=5in, clip=true}
\caption{(a) JQP Current $I_1$ through the left SSET ($V_{1}$ = 435 $\mu $V), plotted in grayscale as a
function of $V_{g1}$ and $V_{2}$ (black indicates no current, while white corresponds to 1.5 pA). The dotted
line indicates the peak
corresponding to $n_{2}$ = 0. (b) Four line traces selected from the data in
(a), offset in proportion to $V_{2}$. }
\label{fig3}
\end{figure}

\section{JQP ENHANCEMENT}
During our experiments in the superconducting state, the left SSET was 
biased to be in the JQP regime.  When 
$V_2$ was grounded, sweeping the gate voltage $V_{g1}$ produced a series
of small peaks ($\sim $0.5 pA) periodic with gate charge of $e$.  
Typically one expects two peaks per period, corresponding to CP 
resonances in either junction.  In our device, the second JQP peak
was barely observable, suggesting that the tunnel barrier
resistances (and thus the Josephson energies) were very dissimilar.

The effects of biasing the right SSET are shown in Figure 3, where $I$ is 
plotted as a function of $V_{g1}$ for various values of $V_{2}$. $V_{g2}$ was counterswept to cancel the cross-capacitance from $V_{g1}$
to the right SSET, fixing the induced charge of the right SSET in each gate
sweep.  Capacitive effects due to the changing asymmetric bias voltage were not
cancelled, however, resulting in a slight leftward shift of the peaks for
increasing $V_{2}$.

For small $V_{2}$, only one (or
two) peaks are visible per period. Beginning at a bias of about 470 $\mu $V,
up to four distinct peaks are seen per period. This bias is too small to allow
successive QP tunneling, so the appearance of extra peaks may indicate a
parallel JQP cycle in the right SSET.  As an example of the complex behavior
seen in this regime, at $V_{2}$ = 570 $\mu $V (Fig. 3(b)), peaks
corresponding to an even $n_2$ are all larger than those with odd $n_2$. The
four-peak structure continues up to 
$V_{2} \sim$ 800 $\mu $V ($\sim 4\Delta/e$). 
At higher biases the peaks disappear one by one, then 
re-emerge with about three times their low-bias amplitude. 
The peak heights continue to increase for even
higher $V_{2}$; no saturation in the peak height was seen in similar
experiments where $V_2$ was swept up to 5 mV.  This enhancement of 
the JQP peaks is in stark contrast to the normal state measurements, where
the peak height only diminished with increasing $V_2$ (Fig. 2(b)).

\section{MODEL AND DISCUSSION}

Although we cannot
presently account for each peak's behavior, it is clear from the 
sharp sidepeaks that each SSET is sensitive to the other's 
charge.  In this section we introduce a possible mechanism for the peak
enhancement in the left SSET above $V_2 \sim 4\Delta/e$, where the current 
through the right SSET is carried predominantly by quasiparticles.

Part of the explanation for the enhancement must take into account
events involving the simultaneous transfer of charge in both SSET's.  The two
processes of interest in our model are depicted in Figure 4, for the case when
a CP can resonantly tunnel into the left SSET only if an extra QP
is resident in the right SSET.  Using the notation ($n_1$,$n_2$) 
to refer to the 
combined charge state having $n_1$ and $n_2$ extra electrons on the two 
respective islands, the states (0,1) and (2,1) are resonant, 
but (0,0) and (2,0) are not.  

\begin{figure}
\epsfig{file=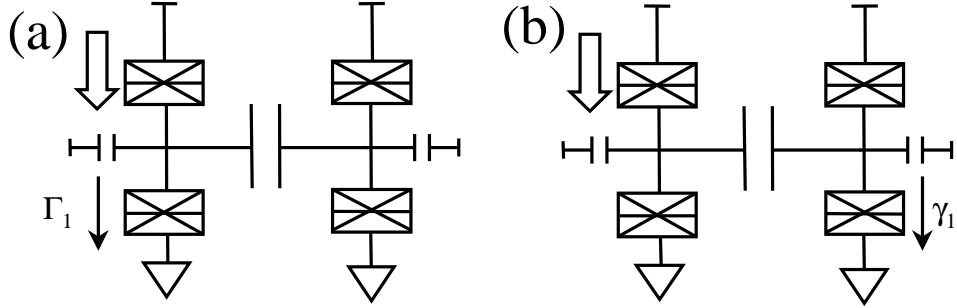, width=5in, clip=true}
\caption{There are two ways for the Cooper pair tunneling resonance to decay:
  (a) the left SSET emits a QP, (b) the right SSET emits a QP.}
\label{fig4}
\end{figure}

In the usual JQP process (Fig. 4(a)), the CP tunnels into the left SSET while
a QP tunnels off with rate $\Gamma_{1}$, so that (2,1) decays to (1,1).  
The cycle is completed when (1,1) decays to (0,1).  
In coupled SSET's, the tunneling of the CP can also be
coincident with a QP decay from the $right$ SSET, so that (2,1) decays to (2,0)
with rate $\gamma_{1}$,
leaving an extra CP in the left SSET.  This CP then spontaneously
decays via two QP decays to ground, and the cycle can
restart after a QP tunnels into the 
right SSET and brings the left SSET back into resonance.

Just as the charge of the right SSET affects the CP resonances in the 
left SSET, the charge of the left SSET affects the QP tunneling rates of 
the right SSET.  The voltage across a junction must exceed $2\Delta / e$ 
for a QP to tunnel across it (ignoring the small subgap conductance).  
Due to the influence of the strong mutual coupling, some of these QP 
tunneling events 
can be blocked depending on the charge of the other island.  
Our model assumes that the right SSET can only undergo QP decay to ground 
when the left SSET has a charge $n_1 \geq 2$ (ie. if a CP is resident), but
a QP can tunnel into the right SSET for $n_1 \leq 2$. The allowed (and disallowed) transitions are summarized in Figure 5(a).  Transitions in 
the left (right) SSET are denoted by $\Gamma_i$ ($\gamma_i$).

Since $E_{c} << 2\Delta$, one
can show that the (allowed) QP rate through each junction, regardless of the 
charge state under consideration, is well-approximated (within a 
factor of $\sim$ 2) by
$\Gamma_{QP} \sim V/eR \sim 2\Delta/e^2R$, where 
$V$ is the voltage across the junction, and $R$ is the normal-state tunneling
resistance.  The $\Gamma_i$ rates all occur through the
lower left junction, so we set:
$\Gamma_1 = \Gamma_2 = \Gamma_3 = \Gamma_4$.  Likewise, the $\gamma_i$ rates
(except $\gamma_1$) all occur through the upper right junction, so 
$\gamma_2 = \gamma_3 = \gamma_4$.  The $\Gamma_i$ rates are fixed, since $V_1$
is fixed, but the $\gamma_i$ rates will increase along with $V_2$.

We will assume that $\hbar\Gamma_{1} < E_{J}$, corresponding to
asymmetric tunnel barrier resistances for the left SSET (as mentioned
in Section 3).  This condition means that the JQP current is 
bottlenecked by slow QP decay.  We will also assume that 
$\gamma_1 > \gamma_2$, corresponding to asymmetric resistances for the right SSET.
The asymmetries are necessary to produce peak enhancement in our model.

Based on these assumptions, we have calculated the peak JQP current as a
function of $\gamma_1$ (Fig. 5(b)), using a density matrix 
formulation\cite{Gurvitz}.  The plot shows a clear enhancement of $I_1$ for 
$\gamma_1 > 0$.  The enhancement is even more pronounced if the 
asymmetry between $\gamma_1$ and $\gamma_2$ is stronger.

\begin{figure}
\epsfig{file=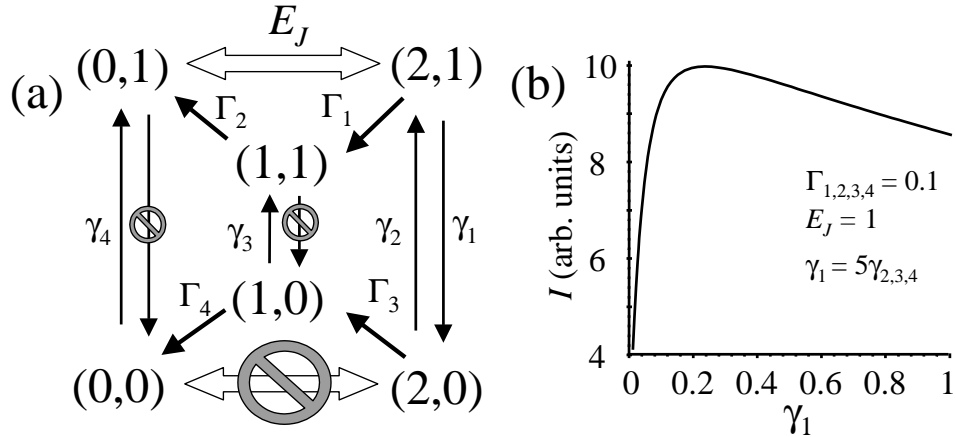, width=5in}
\caption{(a) Schematic of the various tunneling processes allowed in our model.  
The CP can only resonantly tunnel into the left SSET when an extra QP
is present in the right SSET ($n_{2} = 1$), and this QP can only tunnel out
if the CP is present.  (b) Result of the density matrix model, showing current enhancement as $\gamma_{1}$ is increased from zero.}
\label{fig5}
\end{figure}

The enhancement can be understood by considering the effect of $\gamma_1$
on the charge state populations.  The current can be shown
to be equal to:
\begin{equation}
I_1 = 2e(P_{21}\Gamma_1 + P_{20}\Gamma_3) \sim 2e(P_{21} + P_{20})\Gamma_1
\end{equation}
where $P_{ij}$ is the population of the ($i$,$j$) charge state,
corresponding to the appropriate diagonal element of the density
matrix.  The current is directly related to the probability of
finding a CP in the left SSET.  We find that $P_{20}$ increases dramatically with $\gamma_1$,
while $P_{21}$ is only slightly suppressed.  In other words, the left 
SSET becomes {\it more likely} to contain a CP when the right SSET carries
a current of QP's.  Due to the coherent nature of CP tunneling, one
cannot be sure that the CP is on the island until a QP is emitted. 
In the usual JQP process this QP tunnels off the island itself with rate
$\Gamma_1$ (Fig. 4(a)), immediately destroying the CP.  
The larger this rate is, the less likely that the SSET
contains an extra CP since the lifetime of the state is shorter and the
resonance is weakened.  In the coupled-SSET system, however, the 
presence of the CP can also be detected by the emission of a QP with
rate $\gamma_1$ from the other SSET (Fig. 4(b)).  The larger this rate is
(up to a point), the {\it more} likely that the left SSET contains an extra CP,
since this tunneling event does not change the charge of the left SSET.

In conclusion, we have observed a striking modification of the JQP current flowing
through one SSET as a result of its strong Coulomb interaction with another SSET.  
We interpret these results as being a quantum measurement effect, since the
tunneling rates for each SSET are highly sensitive to the other SSET's charge, and thus one SSET can
detect the presence or absence of a Cooper pair in the other SSET.

The authors gratefully acknowledge the input of Caspar van der Wal, K. K. Likharev, and M. Wegewijs.  This research was supported by CHARGE,
Esprit Project No. 22953, and by Stichting voor Fundamenteel
Onderzoek der Materie (FOM).

\end{document}